\documentclass[12pt,final]{article}
\usepackage{graphicx}
\usepackage{epsfig}
\usepackage{dcolumn}
\usepackage{bm}

\usepackage{amssymb,amsmath}
\usepackage{gensymb}
\usepackage{multirow}
\usepackage{booktabs}
\makeatletter
\def\@biblabel#1{(#1)}
\makeatother
\usepackage{setspace}

\usepackage{booktabs,rotating}
\usepackage{fancyhdr}
\usepackage{booktabs,caption}
\usepackage[flushleft]{threeparttable}
\usepackage{enumerate,subfigure,tabularx,longtable}
\usepackage{longtable}
\usepackage[utf8x]{inputenc}
\UseRawInputEncoding
\usepackage{gensymb}
\usepackage {threeparttable} 

\newcommand{\et}{\textit{et al.}}

\newcommand{\comment}[1]{}

\usepackage{lineno}   

\def\gsim {\mbox{\hbox{ \lower-.6ex\hbox{$>$}
\kern-1.12em \lower.5ex\hbox{$\sim$}\kern+.35em}}}
\def\lsim {\mbox{\hbox{ \lower-.6ex\hbox{$<$}
\kern-1.12em \lower.5ex\hbox{$\sim$}\kern+.35em}}}
\makeatletter
\let\@fnsymbol\@arabic
\makeatother

\begin{document}


\title{\vspace{-2.0cm} 
Estimating the disjoining pressure of liquid nanofilms from molecular dynamics simulations via implicit treatment of the bulk liquid phase}

\author{Jianzhou Wang$^{\dag}$, Zufeng Zuo$^{\dag}$, Denvid Lau$^{\ddag}$, Yuzhu Wang$^{\S}$, \\ 
Pinqiang Mo$^{\dag}$, Yukun Ji$^{\dag}$, Liyuan Yu$^{\dag}$, and Yafan Yang$^{\dag,\ddag,\ref{fn:1}*}$ \\
\\[-15pt] 
\small  $^{\dag}$State Key Laboratory of Intelligent Construction and Healthy Operation \\
\small  and Maintenance of Deep Underground Engineering, \\
\small  China University of Mining and Technology, 
\small  Xuzhou, Jiangsu, China. \\
\small $^\ddag$Department of Architecture and Civil Engineering, \\
\small  City University of Hong Kong, 
\small  Hong Kong, China. \\
\small $^\S$Petroleum Engineering Department, College of Petroleum Engineering and Geosciences, \\
\small King Fahd University of Petroleum and Minerals, Dharan, Kingdom of Saudi Arabia.\\
}

\date{\today}
\maketitle

\footnotetext{\label{fn:1}$^*$ To whom correspondence should be addressed, email: yafan.yang@cumt.edu.cn.}

\newpage
\begin{abstract}
The commonly adopted constant bulk liquid density approximation for estimating disjoining pressure in liquid nanofilms, although justified by the low compressibility of liquids, can introduce significant errors in its evaluation. We hypothesize that the bulk liquid density is a thermodynamic state variable that depends on film thickness and local pressure, and should therefore be determined self-consistently to accurately capture interfacial forces under confinement.
A thermodynamically consistent molecular dynamics simulation framework is developed by coupling a surface-tension-based formulation with bulk equations of state obtained from independent simulations. An iterative algorithm is employed to simultaneously determine bulk liquid density, film thickness, and disjoining pressure. The method is applied to water and argon nanofilms, and results are benchmarked against conventional constant-density approaches and alternative computational routes based on chemical potential calculation.
The proposed framework provides a computationally efficient and thermodynamically rigorous route for evaluating disjoining pressure without requiring explicit full coexisting system simulations. Incorporating density variations significantly improves the accuracy and consistency of predicted disjoining pressures. For water nanofilms, a 3.6\% deviation in density can lead to up to 77\% overestimation of disjoining pressure at a thickness of 12~\AA, whereas the effect is weaker but still non-negligible for argon. The proposed framework restores the inverse-cubic scaling with film thickness predicted by Hamaker theory and reduces discrepancies between independent computational methods. Overall, the results demonstrate that self-consistent treatment of bulk thermodynamics is essential for quantitatively reliable evaluation of surface forces in confined fluid systems.
\end{abstract}

\clearpage

\section{Introduction}
Thin liquid films are the fundamental structural elements of foams, emulsions, wetting layers, and confined fluid phases\cite{ivanov2023thin,karakashev2015hydrodynamics}. Their stability, thickness regulation, rupture dynamics, and response to external stimuli such as applied pressure or ionic concentration are governed by the disjoining pressure, which is a thermodynamic potential arising from the net balance of van der Waals, electrostatic, and structural forces across interfaces\cite{derjaguin1974structural,derjaguin1978question}. This potential dictates film equilibrium, metastability, and coalescence resistance, positioning it as a cornerstone of colloid and interface science. Its influence naturally extends to confined fluids and porous media, where it couples directly with capillary pressure to control phase transitions, hysteresis, and stress transmission across nanometric gaps and interparticle spaces\cite{nguyen2020unsaturated,gray2001thermodynamic,pellenq2009simple}. Accurate quantification is therefore essential, not only for rationalising macroscopic colloid stability, but also for understanding nanofluidic transport, capillary-driven deformation, and poromechanical behaviour in both natural and engineered materials.

However, direct experimental characterization remains challenging because the relevant film thickness is often only a few nanometers. Experimental observations further suggest that free liquid nanofilms become unstable below thicknesses of several tens of nanometers due to long-wavelength fluctuations \cite{bhatt2002molecular,macdowell2014disjoining}. Consequently, only limited experimental measurements of surfactant-stabilized free aqueous nanofilms \cite{bergeron1992equilibrium,karraker2002disjoining,stubenrauch2003disjoining} and solid-wetting nanofilms \cite{sharma1993equilibrium,diakova2003thin,huerre2017laplace,zou2021disjoining} have been reported. In this context, molecular simulations offer a distinct advantage, as the finite size of the simulation domain suppresses such fluctuations and enables the investigation of metastable nanofilms that are otherwise difficult to access experimentally.

Simulating a full coexistence system at the molecular level, involving bulk liquid, bulk gas, and free nanofilms, continues to pose challenges. Experimental configurations used to stabilize nanofilms often rely on additional solid substrates, as schematically illustrated for a typical setup in Fig. 1a \cite{stubenrauch2003disjoining}. However, reproducing such conditions within the limited spatial scales accessible to molecular simulations is nontrivial. Beyond interfacial fluctuations, an additional complication arises from the fact that the disjoining pressure may be positive \cite{bhatt2004monte}, negative \cite{bhatt2002molecular,peng2015methodology}, or oscillatory \cite{bergeron1992equilibrium} depending on the fluid composition. Positive disjoining pressures are generally associated with concave liquid interfaces (Fig. 1a), whereas negative values correspond to convex morphologies (Fig. 1b) \cite{bhatt2002molecular}. Consequently, stabilizing coexistence systems typically requires careful tuning of the fluid-solid interaction strength, pore width, and system size, making it difficult to establish robust simulation protocols applicable across different chemical systems. Moreover, explicitly representing all coexisting phases demands substantially larger simulation domains and, therefore, considerably higher computational cost. As a result, to the best of our knowledge, direct molecular simulations of free nanofilm systems composing all coexistence phases have not yet been reported, while only a limited number of full coexistence simulations involving liquid nanofilms wetting a solid substrate are available \cite{fang2019structure,sun2020molecular,li2024hygroscopic}.

Several molecular simulation strategies have been proposed in the literature for evaluating the disjoining pressure of free nanofilms without directly simulating a full coexistence system. Monte Carlo approaches \cite{bhatt2004monte,winter1999computer}, particularly Gibbs ensemble methods, are in principle well suited for coexistence problems because they allow particle exchange between distinct phases through auxiliary simulation boxes. However, for large-scale molecular simulations, the sequential nature of traditional Monte Carlo algorithms inherently restricts their parallel efficiency compared to molecular dynamics (MD), thereby imposing stricter limits on their computational scalability \cite{heffelfinger2000parallel,wang2024multiscale}.

MD approaches for simulating free nanofilm systems typically simplify the system by excluding the explicit bulk liquid phase and considering only the gas-nanofilm configuration (see Fig. 1c) \cite{bhatt2002molecular,peng2015methodology,yang2026resolving}. For example, Bhatt \et \ \cite{bhatt2002molecular,bhatt2003molecular} developed a method based on the Gibbs-Duhem relation to determine the disjoining pressure using chemical potential values calculated from molecular simulations. In addition, when a compatible equation of state is available for the molecular model used in the simulations, the disjoining pressure can be directly determined from the chemical potential-pressure phase diagram \cite{bhatt2002molecular}. However, the applicability of such approaches remains constrained by the availability and accuracy of the equation of state for specific intermolecular force fields. Moreover, these methods rely on evaluating chemical potential in the low-density gas phase, where Widom insertion is generally more reliable than in dense liquids. This requirement significantly limits applicability at low temperatures because the coexistence gas density becomes exceedingly small \cite{bhatt2003molecular}.

These limitations have motivated the adoption of an alternative approach. Specifically, a thermodynamic method has been proposed in which the disjoining pressure is determined from the derivative of the film surface tension with respect to film thickness, based on the Frumkin-Derjaguin theory \cite{peng2015methodology,henderson2005statistical,benet2014disjoining}.  
Subsequent study, incorporating thermodynamically consistent definitions of film thickness and long-range dispersion interactions, has further improved the predictive accuracy of this method \cite{yang2026resolving}. Compared with the chemical-potential-based (CP) method, the surface-tension-based (ST) method provides a more convenient framework for MD simulations of confined nanofilms \cite{peng2016modelling}.

However, because the bulk liquid phase is absent in these MD methods, the thermodynamically consistent Gibbs film thickness is usually evaluated by assuming a constant bulk liquid density \cite{bhatt2002molecular,yang2026resolving,bhatt2003molecular}. This approximation is justified by the low compressibility of liquids, even though the bulk liquid pressure actually varies with film thickness. In the CP method\  \cite{bhatt2002molecular,bhatt2003molecular}, the bulk liquid density mainly influences the determination of film thickness while exerting relatively little effect on the calculated disjoining pressure. In contrast, the ST method is considerably more sensitive to bulk density because film thickness appears explicitly as a differential thermodynamic variable \cite{peng2015methodology,yang2026resolving,peng2016modelling}. Although these approaches have produced reasonable agreements \cite{yang2026resolving}, the thermodynamic implications of neglecting the bulk liquid phase have not been systematically examined.

The validity of the constant-density approximation becomes increasingly questionable under conditions where compressibility effects are non-negligible, particularly at high disjoining pressures. Accurate treatment of the bulk phase may also be important in related confined-fluid systems, such as gas films formed during droplet coalescence, where the thermodynamic state of the surrounding gas phase strongly influences interfacial behavior \cite{aarts2008droplet}. These considerations suggest that rigorous incorporation of bulk-phase thermodynamics is necessary for quantitatively reliable prediction of nanofilm properties.

In this work, we address this limitation by developing a thermodynamically consistent method that implicitly incorporates bulk-phase information into MD results obtained solely from simulations of bulk gas and a nanofilm. Building upon the ST method \cite{peng2015methodology,yang2026resolving}, the proposed method self-consistently couples nanofilm properties with bulk-phase properties through pressure-density relationships determined from separate bulk simulations. The method has been applied to free liquid nanofilms in both water and argon systems. The results demonstrate that rigorous treatment of bulk liquid properties becomes increasingly important under high-disjoining-pressure conditions and substantially improves the accuracy of the prediction of the nanofilm properties. 

By rigorously incorporating bulk-phase thermodynamics without the computational burden of explicit coexistence simulations, the proposed method provides a robust and general MD framework for quantifying disjoining pressures in fluid nanofilms. Its applicability extends to both free-standing and solid-supported films, making it a versatile tool for generating high-fidelity interfacial data across a wide range of colloidal and confined systems\cite{ivanov2023thin,brochard2020revisiting,zhang2026sorption}. Such data are essential for validating continuum models of film stability, wetting, and capillary phenomena, and for establishing quantitative links between molecular surface forces and macroscopic observables such as contact angles, film rupture thresholds, and colloidal interaction potentials. In doing so, this work contributes to a deeper mechanistic understanding of the forces that govern the behaviour of thin liquid films in technological and natural contexts, from emulsion formulation to nanofluidic transport.


\section{Method}

\subsection{ST method}

Following the Frumkin-Derjaguin framework, the disjoining pressure ($\Pi$) can be defined as the negative derivative of the Helmholtz interfacial free energy, $F(h)$, of the coexisting system with respect to the film thickness ($h$) under canonical-ensemble conditions \cite{peng2015methodology,henderson2005statistical,benet2014disjoining}:

\begin{equation}
	\label{eq:1}
	\Pi = -\cfrac{\partial F(h)}{\partial h}.
\end{equation}

For free-standing films with planar fluid-fluid interfaces, $F(h)$ is equivalent to the film tension $\gamma_f'$, and also equal to twice the surface tension $\sigma_f$ (see Fig.~1c) \cite{yang2026resolving}:

\begin{equation}
	\label{eq:2}
	F(h)=\gamma_f'=2\sigma_f.
\end{equation}

The surface tension $\sigma_f$ can be calculated within an MD simulation of the system with a setup shown in Fig. 1c using Bakker's equation \cite{bakker1928kapillaritat,green1960molecular}:
\begin{equation}
	\label{eq:3}
	\sigma_f = \cfrac{\gamma_f'}{2}=\cfrac{1}{2}\int_{-\infty}^{+\infty} \Big[P_{zz}-\cfrac{1}{2}(P_{xx}+P_{yy})\Big] dz,
\end{equation}
where $P_{xx}$, $P_{yy}$, and $P_{zz}$ denote the diagonal entries of the pressure tensor along the $x$, $y$, and $z$ axes, respectively, and the $z$ direction is taken perpendicular to the interface. Note that the ST method is extensible to films wetting solid substrates \cite{benet2014disjoining}; however, the first equality in Eq.~(2) no longer holds, and an alternative method can be used to estimate the fluid-solid interfacial free energy \cite{yang2025estimating,ghoufi2025atomistic,di2025solid}.

Accurate calculation of the film thickness $h$ is critical for determining the disjoining pressure in the ST method, as $h$ appears as a differential variable in Eq.~1.
Different definitions of film thickness have been reported in the literature \cite{bhatt2002molecular,peng2015methodology,bhatt2003molecular,ivanov1975thermodynamics2}. Nevertheless, for a pure system comprising bulk liquid, bulk gas, and liquid-like film phases, the thermodynamically consistent expression of the film thickness is given by \cite{yang2026resolving,henderson2005statistical,benet2014disjoining,ivanov1975thermodynamics2}:
\begin{equation}
	\label{eq:4}
	h=\cfrac{N/A-\rho_vL_z}{\rho_l-\rho_g},
\end{equation}
where $N$ denotes the total number of molecules in the system, $A$ denotes the cross-sectional area, $L_z$ represents the box length along the $z$-direction, and $\rho_l$ and $\rho_v$ are the respective bulk densities of the liquid and gas phases. According to this definition, two dividing surfaces are introduced to describe the two film-gas interfaces, and the film thickness is taken as the distance between these surfaces. In essence, the dividing surfaces split the system into hypothetical fluid regions, each having a uniform density equal to that of the bulk liquid. This thickness is defined as a thermodynamic quantity rather than a direct geometric measurement of the film's physical span \cite{yang2026resolving,henderson2005statistical}. It is intended to act as the variable that is strictly conjugate to the disjoining pressure \cite{henderson2005statistical,ivanov1975thermodynamics2}.

Because the simulation setup (Fig.~1c) excludes an explicit bulk liquid phase, the bulk liquid density $\rho_l$ in Eq.~4 is generally treated as invariant with respect to $h$. In practice, $\rho_l$ is estimated from the density measured in the middle region of the thickest nanofilm considered in the simulations \cite{yang2026resolving,bhatt2003molecular}. This approximation can be reasonable given that the liquid exhibits weak compressibility \cite{bhatt2002molecular,bhatt2003molecular,linstrom2001nist}. As mentioned in the introduction section, the approximation has not been systematically examined. The validity of the constant-density approximation becomes increasingly questionable under certain conditions.

\subsection{Self-consistent treatment of bulk liquid}

A key contribution of this study is incorporating the variable bulk liquid density into the estimation of film thickness $h$ to systematically examine its impact on the disjoining pressure $\Pi$. We propose a thermodynamic framework that treats the bulk liquid phase implicitly. Fig. 2 schematically illustrates both the conceptual physics governing this implicit coupling strategy and its detailed numerical execution through a Picard iteration algorithm.

As schematically illustrated in Fig. 2a, a liquid-like nanofilm bridging two solid substrates coexists with its surrounding bulk gas and bulk liquid phases with a negative disjoining pressure. However, direct MD simulations of such fully explicit macroscopic coexistence systems are prohibitive, as mentioned in the Introduction section. To address this challenge, our methodology decouples the complex three-phase system into a dual-track simulation scheme. In the main molecular simulations, we only explicitly model the isolated film system comprised of the liquid-like nanofilm and the bulk gas phase, as shown in the middle-right panel of Fig. 2a. Crucially, the thermodynamic influence of the omitted bulk liquid phase is not discarded; instead, it is introduced implicitly through a predetermined empirical ``equation of state" (EOS), $\rho_l(P_l)$, obtained from separate bulk-phase simulations (lower-right panel of Fig. 2a). Under this self-consistent coupling framework, the bulk liquid density $\rho_l$ inside the nanofilm is no longer treated as an idealized constant value. Instead, $\rho_l$ becomes a dynamic state variable that responds to variations in the local liquid phase pressure induced by changes in the film thickness $h$.

The numerical convergence of the coupled system is achieved using a Picard iteration algorithm, whose complete computational procedure is detailed in the flowchart of Fig. 2b. At the beginning of the algorithm, a set of independent MD simulations is conducted for film systems characterized by varying total particle numbers $N$. From these simulations, the film surface tension $\sigma_f(N)$, the bulk gas density $\rho_g(N)$, and a preliminary guess for the bulk liquid density $\rho_{l0}$ (taken at the center of the thickest film along the $L_z$ direction) are extracted.

Then, separate bulk gas phase simulations are executed to establish the gas pressure-density relation $P_g(\rho_g)$ based on pre-calculated $\rho_g(N)$ (upper-right panel of Fig. 2a).
After that, separate bulk liquid phase simulations are carried out under different pressures (lower-right panel of Fig. 2a). Note that the pressure range can be selected based on the initial $\Pi$ and $P_g(\rho_g)$ values from earlier simulations. Then, the liquid phase EOS $\rho_l(P_l)$ can be fitted. 

The iteration index is then initialized with $k=1$, and the initial film thickness $h^{(0)}(N)$ for each system is calculated based on the definition (Eq. 4) using the initial constant density assumption $\rho_{l}(N_{max})$. At each iteration step, the current data pairs of film thickness and surface tension, $(h^{k-1}, \sigma_f)$, are fitted to an exponential function \cite{li2024hygroscopic,yang2026resolving,peng2017surface}:
\begin{equation}
	\label{eq:8}
	\sigma_f=a \cdot e^{b\cdot h}+c,
\end{equation}
where $a$, $b$, and $c$ are the fitting parameters in units of $mN/m$, $\textup{\AA}^{-1}$, and $mN/m$. 

According to Eq. 1, the analytical derivative of this fitted curve is taken to yield the current formulation of the disjoining pressure, $\Pi^{k}(h^{k-1})$. Leveraging the mechanical balance across the interfaces, the corrected bulk liquid pressure is computed via $P_l^{k}(h^{k-1}) = P_g(N) - \Pi^{k}(h^{k-1})$. This updated pressure $P_l^{k}$ is subsequently substituted into the pre-calibrated liquid EOS to back-calculate the bulk liquid density $\rho_l^{k}(P_l)$. The newly updated variable liquid density $\rho_l^{k}$ is then fed back into the expression (Eq. 4) to update the film thickness to a more precise, compressible-corrected value $h^{k}(N)$. Finally, the convergence behavior is evaluated by checking whether the maximum relative error of the calculated film thickness between two consecutive iteration steps satisfies the criterion where the maximum relative error of the calculated film thickness between two consecutive iteration steps is less than or equal to $10^{-5}$. If this criterion is not satisfied, the iteration index is advanced, and the algorithm loops back to repeat the surface tension fitting and disjoining pressure calculation. If the condition is successfully met, the iteration terminates, and the final outputs $h$ and $\Pi$ are extracted as the true equilibrium states where the stress state of the nanofilm and the bulk-phase compressibility are simultaneously satisfied.

By employing this implicit coupling iteration, the proposed method significantly enhances the accuracy of property predictions under high negative $\Pi$, where nanofilms are extremely thin and liquid compressibility effects become prominent, all while bypassing the prohibitive computational overhead and physical complexities inherent in explicit multi-phase MD simulations.

\subsection{Simulation Details}

The simulation details are consistent with our previous study \cite{yang2026resolving}.
All molecular dynamics simulations were performed in LAMMPS \cite{thompson2022lammps} using the velocity Verlet integrator under full periodic boundary conditions. Water was modeled using the flexible SPC/E potential to accurately capture surface tension effects \cite{lopez2008effect,berendsen1987missing,buyukozturk2011structural}, while argon parameters were sourced from Sherwood and Prausnitz \cite{sherwood1964intermolecular}. 
The long-range Lennard-Jones (LJ) and Coulombic interactions are considered utilizing the particle-particle-particle-mesh (PPPM) solver (relative tolerance of 10$^{-4}$). 

The film system results are obtained directly from our previous simulations \cite{yang2026resolving}. The film domain geometry replicates the configurations used in the CP method \cite{bhatt2002molecular,bhatt2003molecular} to ensure direct comparability. The cross-sectional areas ($x \times y$) are set to $49.056~\text{\AA} \times 49.056~\text{\AA}$ for argon and $36.000~\text{\AA} \times 36.000~\text{\AA}$ for water, with the box length in the $z$-direction equal to three times the lateral dimension.
For bulk gas simulations, we use cubic boxes with a side length of 100 \text{\AA}. The bulk gas systems are simulated at a fixed density corresponding to the vapor phase density obtained from the film simulations. In contrast, for bulk liquid simulations, we employ cubic boxes whose cross-sectional areas match those of the film systems, but with the zz-axis length set equal to the lateral width. 

Both the film systems and bulk gas simulations are performed in the canonical ($NVT$) ensemble, with temperature controlled using a Nos\'e-Hoover thermostat. Meanwhile, the bulk liquid simulations are carried out in the isothermal-isobaric ($NPT$) ensemble, with both temperature and pressure regulated using a Nos\'e-Hoover thermostat and barostat.
Timesteps of 5~fs and 1~fs were applied to the argon and water systems, respectively. Argon simulations ran for a 7.5~ns equilibration stage followed by a 22.5~ns production phase, whereas water systems were equilibrated for 3~ns and sampled for 6~ns. Finally, production data were segmented into three independent blocks to quantify statistical uncertainties.

\section{Results and Discussion}

\subsection{Pure water system}

We first applied the proposed method to the pure water system. The corresponding film systems with various $N$ have been simulated in our previous work \cite{yang2026resolving}, from which the surface tension data are readily available. For each film system, we obtained the vapor-phase density in the bulk gas region $\rho_g$. We found that $\rho_g$ ranges from 1.086e-4 to 1.252e-4 $\text{\AA}^{-3}$, and no clear relation between $h$ and $\rho_g$ can be identified due to large uncertainties. Notably, a previous study by Bhatt et al. \cite{bhatt2003molecular} reported a clear increasing trend of $\rho_g$ with $h$ using a shorter simulation time and a similar system size compared to ours. Nevertheless, we find that the variation in $\rho_g$ does not significantly affect the gas pressure $P_g$, as demonstrated by computing the bulk pressures at the lower and upper bounds of the gas density. For instance, the corresponding bulk gas pressure varies from $0.59$ to $0.68$ MPa, a difference of only $0.09$ MPa. This is also consistent with experiment data for water vapor \cite{linstrom2001nist}. Since this difference is negligibly small compared to the disjoining pressure, which is on the order of $\sim10$ MPa, we used an averaged gas density to compute the bulk gas pressure for all $h$. The resulting bulk gas density and pressure are 1.169e-4 $\text{\AA}^{-3}$ and $0.63$ MPa, respectively.

Fig. 3 presents the simulated liquid water densities as a function of pressure at 479.00 K, along with the corresponding fitted curve. Our simulated density data agree well with experimental data \cite{linstrom2001nist}. For example, the experimental water density varies from 0.028673 to 0.031029 $\text{\AA}^{-3}$ as the pressure changes from the saturation point to around 120 MPa.
To obtain the EOS, an empirical formula was employed to fit the data:
\begin{equation}
	\rho_l=C_1 \times P_l^3 + C_2 \times P_l^2 + C_3 \times P_l + C_4,
\end{equation}
where C1-C4 are the fitting parameters. The relation resembles the virial equation of state \cite{schultz2022virial}. However, we do not aim to develop a set of robust parameters applicable over a wide range of temperature and pressure conditions, nor to represent thermodynamic properties beyond the pressure-density relationship. Therefore, the only pressure density data at a specific temperature are used for fitting.
It should be noted that this approach may lead to overfitting, and the resulting parameters are not expected to perform well outside the fitted conditions. Nevertheless, as long as the pressure range remains within that relevant to the film system, and the accuracy of the relation is sufficient for our application, this limitation is acceptable.
A more robust equation of state for fluids at elevated temperature and pressure can be constructed using more sophisticated models \cite{kontogeorgis2006ten,muller2001molecular,duan2025molecular}. However, such approaches typically require simulations over a much broader range of thermodynamic states, leading to significantly higher computational cost. 

Using the bulk gas properties and the EOS for the bulk liquid phase, we performed the self-consistent method. Fig. 4 shows the convergence performance of the method for both bulk and interfacial properties. These values vary significantly during the first three iterations and converge quickly. The iterations terminate at step~6. Additionally, it is found that for films with smaller $h$, the variations of the studied properties are larger.

Table 1 summarizes the bulk liquid and interfacial properties at the initial ($k=0$) and final ($k=6$) iterations for ten water film systems at 470~K. The surface tension $\sigma_f$ is taken from Ref. \cite{yang2026resolving} and increases from 25.92 to 37.42~mN/m as the film thickness increases. The self-consistent iteration reduces the bulk liquid pressure $P_l$ and density $\rho_l$ for systems with relatively small $h$ ($e.g.$, Film 1-4), where the initial pressure is overestimated. 
For example, in Film 1 (the thinnest film), $P_l$ decreases by 15.4\% (from 126.65 to 107.09~MPa) and $\rho_l$ by 1.03\% (from 0.032126 to 0.031795~\AA$^{-3}$). 
For larger $h$ (Film 5-10), the changes between initial and final values become progressively smaller, and for Film~9 and~10, the values are practically unchanged. 
Similarly, the film thickness $h$ decreases slightly after iteration, with the largest relative reduction occurring for the thinnest film.
These results indicate that the self-consistent correction is most significant for strongly confined films, while for thick films where the liquid phase approaches bulk behavior, the initial estimates are already accurate.

Fig. 5a compares the thickness-dependent surface tension values with thickness obtained from the ST method assuming a constant liquid density ($k=0$) against those from the self-consistent method, which implicitly accounts for the bulk liquid density. As expected, the overall agreement between the two fitting curves improves at large $h$, where the simulated surface tension converges to $37.42\,\text{mN/m}$. This value shows a minor discrepancy of only $2.97\%$ compared to the experimental value of $36.34\,\text{mN/m}$ \cite{linstrom2001nist}. Conversely, a clear discrepancy emerges between the two curves at small $h$. Specifically, the fitting coefficient $a$ in Eq. 5 (given in Fig. 5a) is reduced by a factor of 4.5 once the variation in bulk liquid density is taken into account.

Fig.~5b compares the corresponding disjoining pressures with fitting curves. The prefactors of the exponential terms exhibit a significant difference: the prefactor obtained using the constant-density ST method is approximately 5.3 times larger than that of the self-consistent method with corrected liquid density. Accounting for the variation in liquid density generally attenuates the magnitude of $\Pi$ (i.e., yields smaller absolute values). This discrepancy in $\Pi$ becomes more pronounced at smaller film thicknesses. For instance, at $12\,\text{\AA}$, the constant-density method overestimates the magnitude of $\Pi$ by $77.0\%$, whereas at $17\,\text{\AA}$, the overestimation drops to $11.4\%$. For $h > 18\,\text{\AA}$, the difference between the two methods becomes negligible.

We further compared the results of the ST method with those of the CP method. Previously, Bhatt et al. \cite{bhatt2003molecular} implemented the CP method by assuming a constant liquid density when calculating both $h$ and $\Pi$. In this work, we modify the calculation of $h$ to account for the corrected liquid density using our fitted EOS. Note that the Gibbs-Duhem relation used in the CP method also assumes a constant liquid density. This assumption was left uncorrected because it has a relatively minor influence on the calculated $\Pi$ as demonstrated by Bhatt \et \ \cite{bhatt2002molecular}.
Consequently, the overall agreement between the CP and ST methods improves significantly when using the corrected liquid density; the absolute average difference (AAD) between the two methods decreases from $25.07\%$ to $14.69\%$ upon incorporating the bulk liquid density variation.

According to the classical Hamaker theory, the disjoining pressure is expressed as  \cite{israelachvili2011intermolecular}:
\begin{equation}
	\label{eq:10}
	\Pi = -\frac{A_H}{6\pi h^3},
\end{equation}
where $A_H$ denotes the Hamaker constant. To evaluate this theoretical relationship, Fig.~5c plots $\Pi$ against $1/h^3$ alongside the corresponding linear regressions modeled after Eq.~\eqref{eq:10}. The resulting fitted Hamaker constants are explicitly detailed within the figure. 

Evidently, the disjoining pressures computed via the density-corrected methods demonstrate superior conformance to the linear scaling dictated by the Hamaker theory than those calculated under the constant-density assumption. Conversely, treating the liquid density as constant yields artificially elevated Hamaker constants. Furthermore, accounting for the variation in bulk liquid density remarkably reconciles the divergence between the CP and ST methods. Under the constant-density condition, the ST method overestimates $A_H$ by $49.1\%$; however, this discrepancy is suppressed to a mere $18.5\%$ once density corrections are implemented.

\subsection{Pure argon system}

Subsequently, the proposed self-consistent iteration method was applied to the pure argon system. 
The interfacial and bulk properties of argon nanofilms at 100.05~K and 105.93~K, evaluated under the conventional assumption of a constant bulk liquid density, have been documented in our previous work~\cite{yang2026resolving}. 
Consistent with the observations in the pure water system, no discernible trend was found for the bulk gas density as a function of film thickness due to large uncertainties. 
Conversely, Bhatt \et~\cite{bhatt2002molecular} reported a distinct decreasing trend in bulk gas density with increasing film thickness using a shorter simulation duration but a comparable system size. 
Given the negligible variance observed, the dependence of bulk gas density on film thickness was neglected, and averages of bulk gas densities were adopted for the self-consistent framework. 
Specifically, the averaged bulk gas densities (and corresponding vapor pressures) were determined to be $2.40746 \times 10^{-4}$~\AA$^{-3}$ (0.31~MPa) at 100.05~K and $3.63046 \times 10^{-4}$~\AA$^{-3}$ (0.46~MPa) at 105.93~K. 
The uncertainty stemming from these density estimations introduces an error of approximately 0.06~MPa in the bulk gas pressure. 
This margin is orders of magnitude smaller than the calculated disjoining pressures and thus exerts a negligible influence on the final iterative results.

Before implementing the self-consistent framework, the bulk liquid EOS for the argon system must be established to correlate bulk liquid density with pressure. Fig.~6 illustrates the liquid argon density as a function of pressure at 100.05~K and 105.93~K. It is noted that the molecular dynamics simulations systematically underestimate the experimental liquid density by approximately 10\%, compared with experimental data \cite{linstrom2001nist}. This discrepancy is primarily attributed to the intrinsic limitations of the adopted force field parameters for argon. 
To mathematically incorporate these pressure effects into our iterative algorithm, the simulation data points were fitted. The resulting empirical EOS curves, represented by the solid lines in Fig.~6, provide an accurate continuous description of the compressible bulk liquid phase.

Utilizing the determined bulk gas properties and the fitted bulk liquid EOS, the self-consistent iteration framework was executed for the pure argon system. Fig.~7 illustrates the convergence performance of the iterative algorithm for key thermodynamic and structural properties. The calculated variables exhibit exceptionally stable behavior, displaying only minor variations across the successive iteration steps. For all simulated cases across both target temperatures (100.05~K and 105.93~K), the algorithm demonstrates rapid numerical convergence, successfully satisfying the predefined tolerance criteria and terminating by the 3rd or 4th iteration step. 

Consequently, the self-consistent iteration yields generally subtle adjustments to the bulk liquid and interfacial properties, which are quantitatively documented in Table~1. This quick convergence behavior indicates that the initial estimations derived from the simulation configuration are closely aligned with the thermodynamic equilibrium state of the argon nanofilms, requiring only fine-grained iterative corrections to achieve full self-consistency.

Figs.~8a and 8d compare the film thickness-dependent $\sigma_f$ profiles calculated from the ST method assuming a constant bulk liquid density ($k=0$) against those refined via the proposed self-consistent framework. As expected, the self-consistent method does not induce a substantial deviation in the $\sigma_f$-$h$ relationship, primarily because the adjustments made to the absolute film thickness values are relatively minor.  Nevertheless, it is observed that the self-consistent method exerts a more pronounced impact at the higher temperature (105.93~K) than at the lower temperature (100.05~K).
For reference, the experimental surface tension values of pure argon at 100.05~K and 105.93~K are reported to be 9.47~mN/m and 7.28~mN/m \cite{linstrom2001nist}, respectively. At the largest simulated film thicknesses, the molecular dynamics simulations overestimate these experimental values by approximately 20\%. This discrepancy is a recognized characteristic of the underlying intermolecular potential model. Nevertheless, our calculated surface tension values demonstrate excellent agreement with the comprehensive literature data summarized by Werth \et~\cite{werth2013influence}.

Figs.~8b and 8e display the corresponding disjoining pressure profiles as a function of film thickness ($\Pi$--$h$). As expected, the self-consistent method yields $\Pi$--$h$ curves that closely resemble those obtained from the original ST method. At the higher temperature, the difference is larger at small $h$ values.
Note that in the work of Bhatt \et~\cite{bhatt2002molecular}, an EOS was integrated with the CP method, wherein the bulk liquid densities are directly accessible from the EOS, so the thickness corrections are unnecessary. 
Accounting for the thickness-dependent variations in bulk liquid density leads to contrasting shifts in the deviations between the ST and CP methods at different temperatures. Specifically, at 100.05~K, the AAD between the ST and CP methods decreases from 22.0\% to 18.7\% upon implementing the self-consistent framework. Conversely, at 105.93~K, the AAD between the two approaches increases from 7.7\% to 13.5\% when the density variations are taken into account.

Figs.~8c and 8f present the corresponding $\Pi$ profiles as a function of $h^{-3}$. Although the adjustments of properties introduced by accounting for the thickness-dependent bulk liquid density are generally moderate, they significantly refine the extracted Hamaker constants, yielding much closer agreement with the reference CP method. Specifically, at 100.05~K, the overestimation of the Hamaker constant by the ST method relative to the CP method decreases from 23.6\% to 15.7\% when the density-corrected framework is implemented. A more pronounced improvement is observed at 105.93~K, where the overestimation drops from 36.6\% to 18.7\% upon incorporating the bulk liquid density variations. These results underscore that even subtle density corrections are crucial for accurately capturing the true $1/h^3$ scaling law of confined nanofilms.


Overall, the proposed self-consistent framework demonstrates significantly improved accuracy over conventional ST methods that assume a constant bulk liquid density across different film thicknesses. Although the conventional assumption is often considered reasonable due to the weak compressibility of liquids, the present results show that even relatively small variations in liquid density can substantially influence the calculated disjoining pressure, particularly for ultrathin nanofilms. In the water system, the bulk liquid density varies by only approximately 8\% over the investigated pressure range; nevertheless, this variation leads to pronounced differences in the predicted disjoining pressures because the film thickness enters the ST method as a differential thermodynamic variable. For example, at a film thickness of approximately 12~\AA, the constant-density assumption overestimates the magnitude of the disjoining pressure by as much as 77\%, whereas the discrepancy becomes negligible for thicker films ($h > 18$~\AA). These results indicate that accurate treatment of the bulk liquid compressibility is essential under high-disjoining-pressure conditions, where small deviations in film thickness can propagate into large errors in $\Pi$.

The influence of the density correction is significantly more pronounced for water than for argon. This difference mainly arises because water nanofilms exhibit much larger disjoining pressures and stronger confinement effects, making the thermodynamic inconsistency introduced by the constant-density assumption more severe. In contrast, the argon systems show only modest variations after applying the self-consistent correction, and the iterative procedure converges rapidly with only small adjustments to the bulk and interfacial properties. Nevertheless, the correction becomes increasingly important at higher temperatures, as observed for argon at 105.93~K, where the deviations between the corrected and uncorrected ST results become more noticeable at small film thicknesses. Although the overall changes in $\Pi$ remain moderate for argon, the density correction has a significant impact on the extracted Hamaker constants. In both water and argon systems, incorporating the variable bulk liquid density improves the linear scaling between $\Pi$ and $1/h^3$ predicted by the classical Hamaker theory and reduces the discrepancy between the ST and CP methods. These findings suggest that rigorous treatment of bulk-phase thermodynamics is particularly important for obtaining quantitatively reliable disjoining pressures and Hamaker constants in highly confined nanofilm systems.

%
Beyond the specific water and argon systems investigated in this study, the proposed framework provides a general thermodynamic strategy for evaluating disjoining pressures in confined fluid systems without requiring computationally prohibitive and physically complex explicit coexistence simulations.
Because the methodology is fundamentally based on interfacial thermodynamics \cite{henderson2005statistical}, it can be naturally extended to more complex systems, including liquid films wetting solid substrates and confined gas films relevant to droplet coalescence and interfacial transport phenomena. The present work,  therefore, establishes a foundation for future investigations of fluid-solid nanofilm systems commonly encountered in nanoporous materials, shale reservoirs, cementitious materials, and membrane technologies. 
Moreover, although the current validation primarily focuses on comparing the simulated bulk-limit properties, such as liquid density and surface tension, with available experimental data, future work should further assess the predicted disjoining pressures directly against experimental measurements whenever reliable nanoscale data become available \cite{karraker2002disjoining,stubenrauch2003disjoining,diakova2003thin,huerre2017laplace,zou2021disjoining}.
Such developments would further strengthen the applicability of the proposed framework for quantitatively bridging molecular-scale interfacial forces with continuum-scale transport and poromechanical behavior.

\section{Conclusion}

As illustrated in Fig. 9, we developed a self-consistent framework to evaluate the disjoining pressure of nanofilms by coupling the bulk liquid, film thickness, and surface forces. This framework shows that the influence of density variations becomes more pronounced at smaller film thicknesses and is strongly system-dependent. For water nanofilms, this effect is particularly significant: a modest 3.6\% deviation in liquid density leads to a 77.0\% overestimation of the disjoining pressure at 12~\AA. For argon nanofilms, the influence of density variation is comparatively weaker, yet remains non-negligible for ensuring thermodynamic consistency. Importantly, the proposed framework not only reproduces the inverse-cubic scaling of disjoining pressure with film thickness predicted by classical Hamaker theory, but also substantially improves the internal consistency across different computational routes.

The central conceptual advance stems from recognizing a key limitation of the constant-density approximation. Although often justified by the low compressibility of liquids, this assumption can introduce substantial errors in the predicted disjoining pressure. In reality, the bulk liquid density is not fixed but a thermodynamic state variable that depends on both film thickness and the local pressure conditions. Accurately capturing surface forces under confinement, therefore, requires determining the density self-consistently. This perspective leads naturally to a revised framework in which density variations are intrinsically coupled to the disjoining pressure. The primary innovation is the incorporation of this density self-consistency into a surface-tension-based formulation, establishing a rigorous thermodynamic link among bulk compressibility, film geometry, and surface forces. As a result, the framework offers a clearer, more physically grounded description of the nonlinear coupling among thermodynamic variables in nanoconfined films, thereby strengthening the theoretical foundation of thin-film interfacial thermodynamics.

The newly developed method is particularly valuable given the experimental challenges in measuring disjoining pressure, which arise primarily from thermal fluctuations at the nanoscale \cite{bhatt2002molecular,macdowell2014disjoining}. To the best of our knowledge, no experimental data are available for the simple pure systems considered in this study, and existing measurements are mainly limited to stabilized films, such as those involving wetting on solid substrates \cite{sharma1993equilibrium,diakova2003thin,huerre2017laplace,zou2021disjoining} or surfactant-laden interfaces \cite{bergeron1992equilibrium,karraker2002disjoining,stubenrauch2003disjoining}. Consequently, we validated our simulations against properties including bulk density and surface tension. The reasonable agreement with experimental data, as discussed above, supports the reliability of the approach and justifies its extrapolation to predict disjoining pressures at small film thicknesses. 

The experimental difficulty has motivated the development of MD approaches \cite{bhatt2002molecular,bhatt2004monte,peng2015methodology,yang2026resolving}. However, as noted in the Introduction, direct MD simulations of systems comprising coexisting bulk liquid, bulk gas, and free-standing nanofilms remain both computationally demanding and physically complex. Most existing MD studies, therefore, consider only a nanofilm in equilibrium with a gas phase, thereby neglecting the role of the bulk liquid reservoir \cite{bhatt2002molecular,bhatt2004monte,peng2015methodology,yang2026resolving}. In contrast, the present method implicitly incorporates the bulk liquid phase, enabling a more rigorous and thermodynamically consistent treatment. Our results demonstrate that accounting for variations in bulk liquid density leads to significant differences in the predicted disjoining pressure and Hamaker constants compared with the conventional constant-density approach \cite{peng2015methodology,yang2026resolving,bhatt2003molecular}. To further assess the validity of the framework, we benchmarked it against an alternative method based on chemical potential calculations conducted at relatively high temperatures, which is known to exhibit reduced accuracy under low-temperature conditions \cite{bhatt2002molecular,bhatt2003molecular}. Notably, the two independent approaches show substantially improved agreement in both the disjoining pressure and the extracted Hamaker constants once the density variations are properly incorporated.

Future efforts should focus on extending the present framework and further validating its applicability. The method developed in this work holds considerable promise for broader application in the study of interfacial phenomena in confined systems. In particular, extending the approach to more complex multicomponent systems would enhance its relevance to realistic interfacial environments, including liquid- and vapor-like nanofilms interacting with gas, oil, electrolyte, and surfactant-containing phases. Additionally, adapting the framework to describe fluid behavior on solid substrates would provide deeper insight into wetting, spreading, and interfacial stability, thereby expanding its applicability to a wide range of practical and technologically relevant systems.

A synergistic integration of experiment, simulation, and theory is expected to play a pivotal role in validating the proposed method and advancing the understanding of nanofilm thermodynamics. The high-fidelity data generated in this work could help stimulate the development of novel experimental techniques. At the same time, direct comparison with existing experimental measurements of disjoining pressure is essential for assessing the predictive capability of the model and establishing its quantitative reliability. Finally, integration with advanced theoretical frameworks, including classical density functional theory, would provide a systematic approach for describing strongly inhomogeneous systems and for bridging molecular-scale simulations with continuum-level theoretical descriptions.


\bigskip
\bigskip
{\bf{ACKNOWLEDGMENTS\\[1ex]}}
The research is supported by the Fundamental Research Funds for the Central Universities (2025QN1175).

\bibliography{DisP2}


\clearpage

\begin{figure}[tb]
	\centering
	\includegraphics[width=\textwidth]{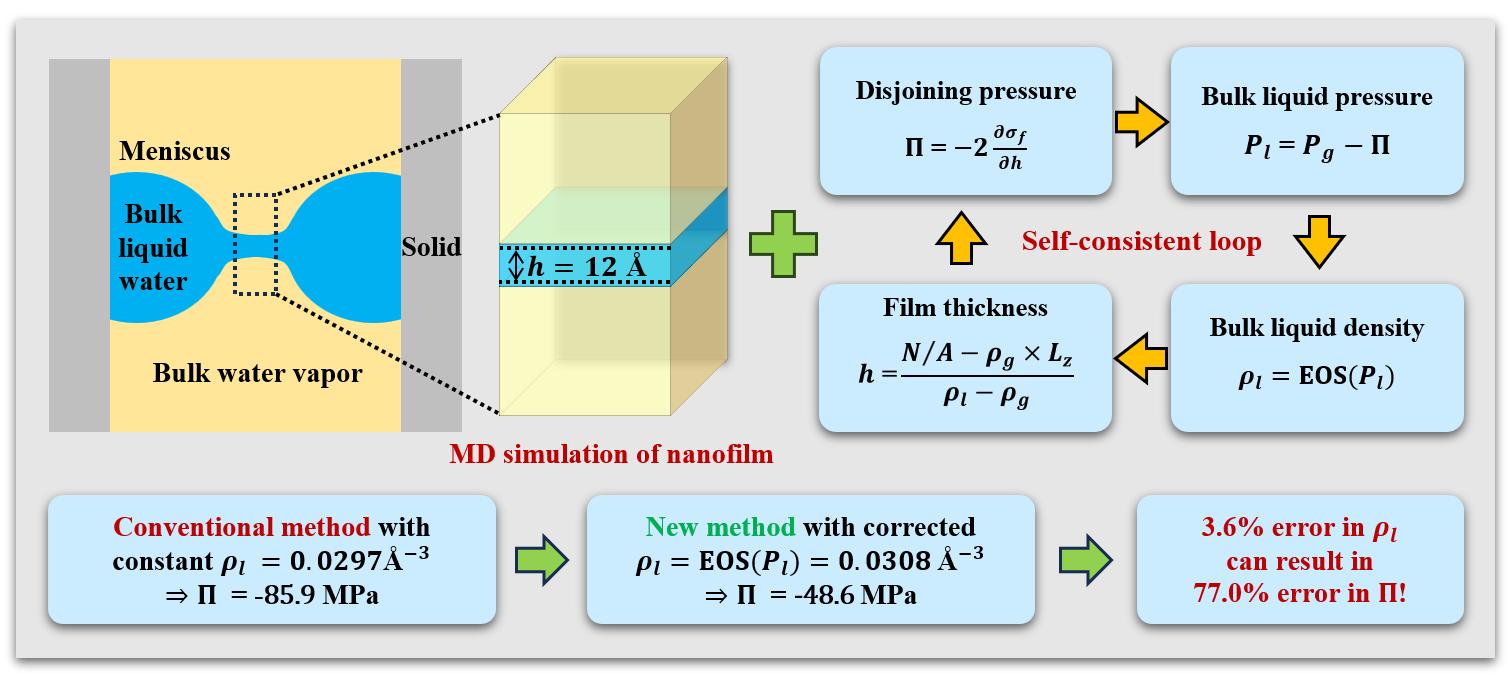}
	
	\vspace{0.5cm}
	
	{\large\bfseries Graphical Abstract\par}
	
	\vspace{1.5ex}
	
	\begin{flushleft}
		Self-consistent determination of disjoining pressure in fluid nanofilms.
		The proposed framework shows that a small error (3.6\%) in the assumed liquid water density can propagate into a large error (77.0\%) in the calculated disjoining pressure at the same film thickness. (This figure is intended for color reproduction on the Web and in black-and-white in print.)
	\end{flushleft}
	
\end{figure}

\clearpage
\begin{figure}[tb]
	\begin{centering}
		\includegraphics[width=0.7\textwidth]{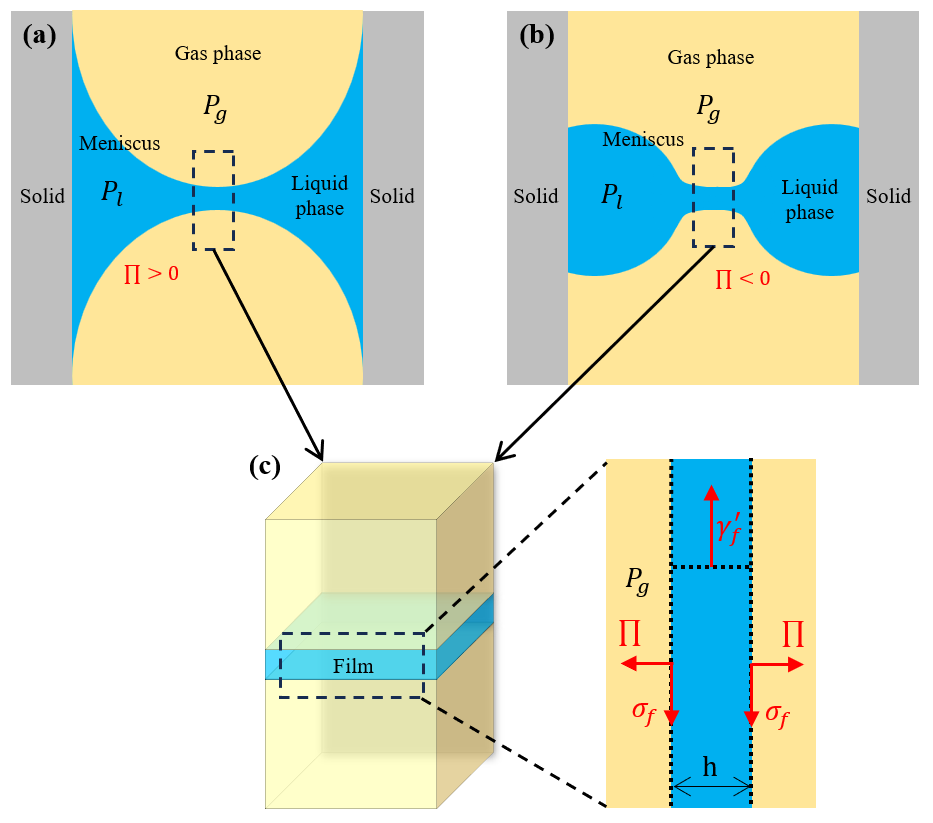}
		\caption{Schematic illustration of liquid nanofilm formation under different signs of disjoining pressure: (a) nanofilm with positive disjoining pressure and (b) nanofilm with negative disjoining pressure. (c) molecular dynamics simulation setup for simulating the equilibrium between the liquid film and bulk gas phase, including the force distribution within the nanofilm. (This figure is intended for color reproduction on the Web and in black-and-white in print.)
		}
		\label{fig:z1}
	\end{centering}
\end{figure}

\clearpage
\begin{figure}[tb]
	\begin{centering}
		\includegraphics[width=0.9\textwidth]{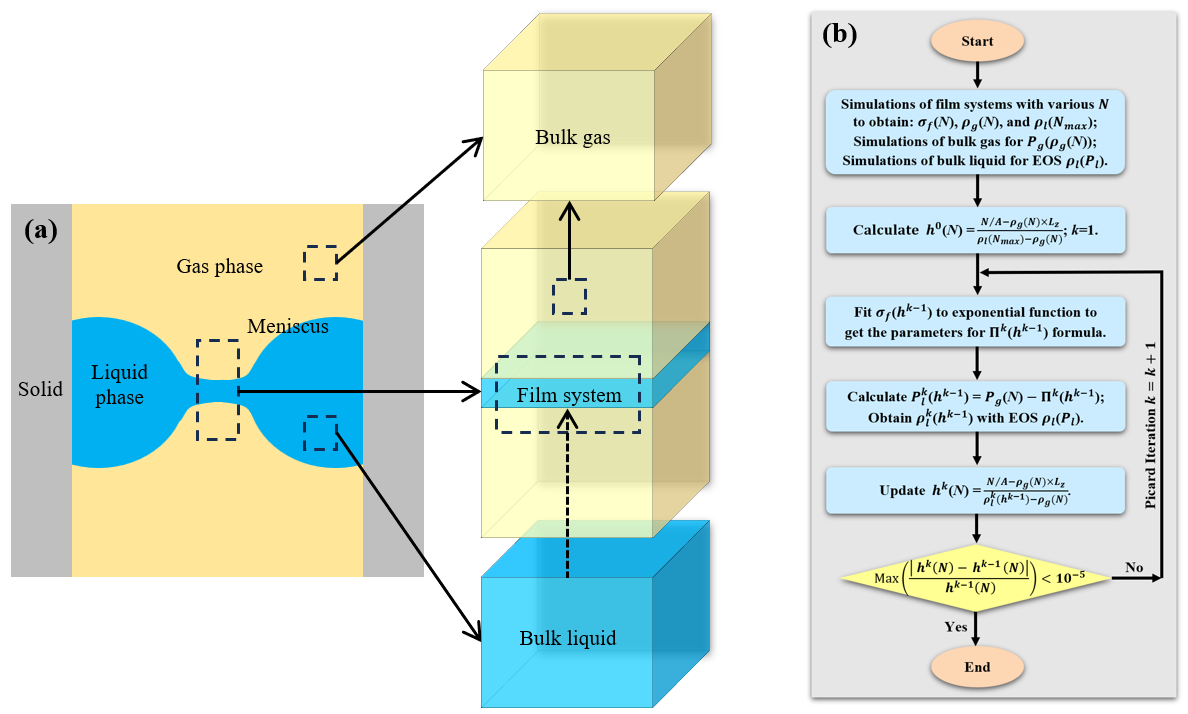}
		\caption{(a) Configuration of bulk phases (upper right and lower right) and the film system (middle right) as illustrated in the schematic nanoconfined system (left) with negative disjoining pressure.
			(b) Flowchart of the iterative algorithm for calculating equilibrium film thickness and disjoining pressure. (This figure is intended for color reproduction on the Web and in black-and-white in print.)
		}
		\label{fig:z2}
	\end{centering}
\end{figure}

\clearpage
\begin{figure}[tb]
	\begin{centering}
		\includegraphics[width=0.7\textwidth]{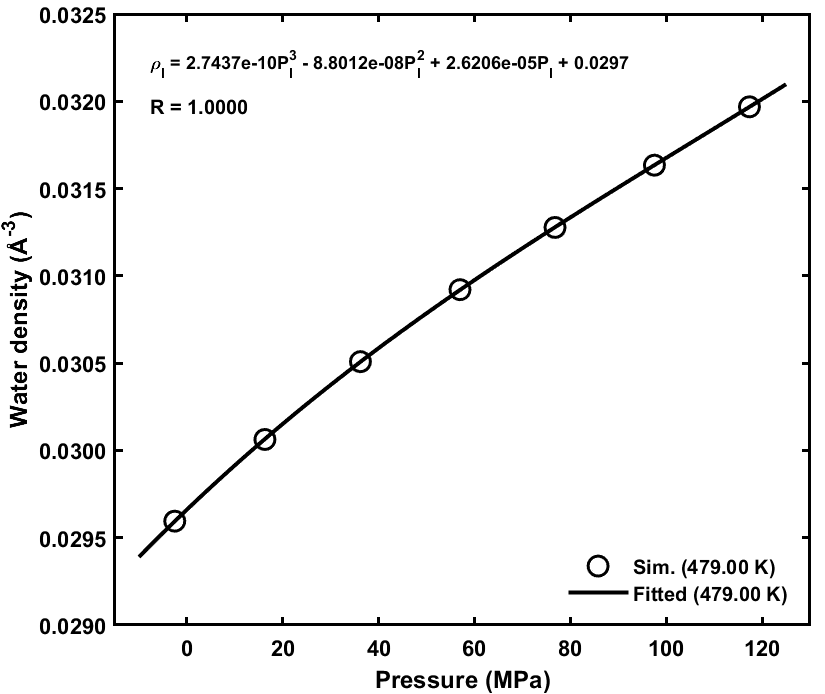}
		\caption{Liquid water density as a function of pressure at 479.00 K. The solid line represents the fitted empirical EOS. (This figure is intended for color reproduction on the Web and in black-and-white in print.)
		}
		\label{fig:z3}
	\end{centering}
\end{figure}

\clearpage
\begin{figure}[tb]
	\begin{centering}
		\includegraphics[width=0.9\textwidth]{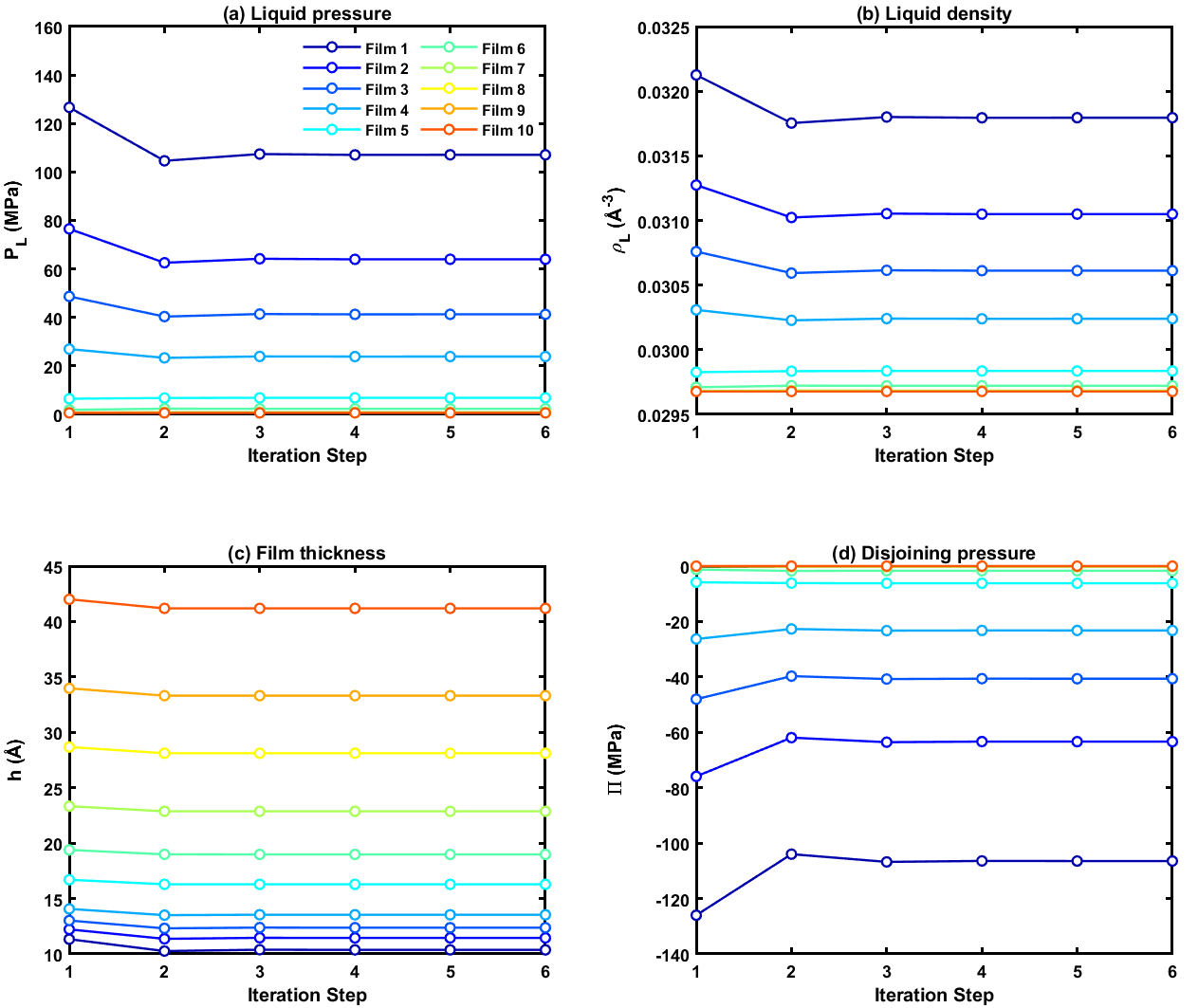}
		\caption{Convergence performance of the iteration algorithm for various properties of the water nanofilm system: (a) liquid pressure, (b) liquid density, (c) film thickness, and (d) disjoining pressure as a function of iteration steps. The labels Films 1-10 correspond to simulated film systems given in Table 1. (This figure is intended for color reproduction on the Web and in print.)
		}
		\label{fig:z4}
	\end{centering}
\end{figure}

\clearpage
\begin{figure}[tb]
	\begin{centering}
		\includegraphics[width=0.5\textwidth]{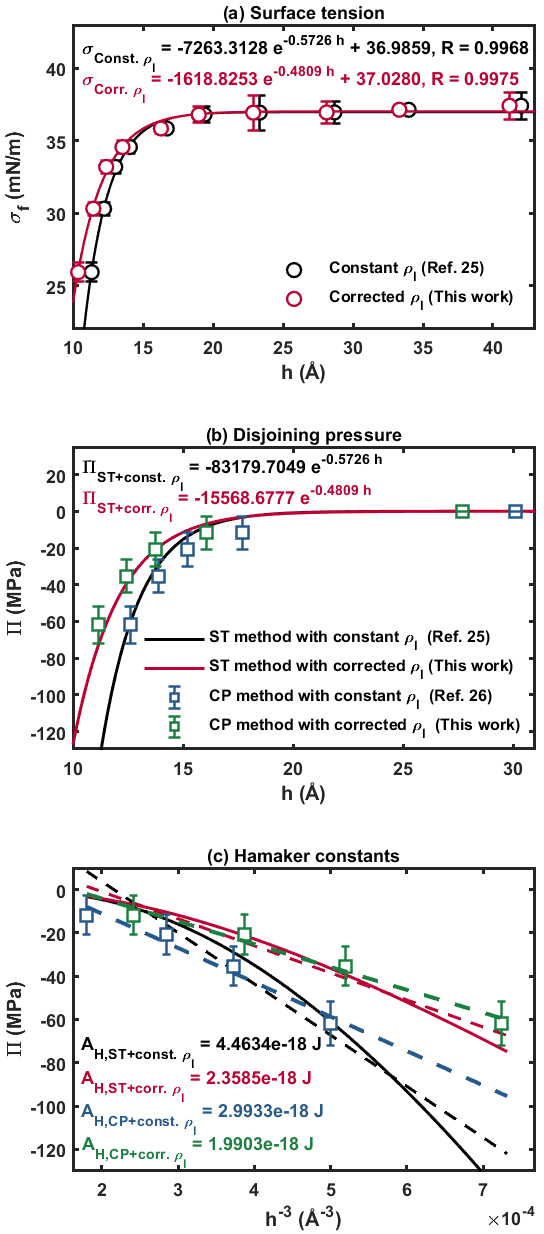}
		\caption{(a) Surface tension of the water nanofilm as a function of film thickness at 479.00 K. Solid lines represent the fitted curves. 
			(b) Corresponding disjoining pressure as a function of film thickness.
			(c) Corresponding disjoining pressure as a function of the inverse cubic film thickness. 
			Dashed lines represent linear fits.
			Our results are compared with those reported by Bhatt \textit{et al.} \cite{bhatt2003molecular} and Yang \textit{et al.} \cite{yang2026resolving}. (This figure is intended for color reproduction on the Web and in print.)
		}
		\label{fig:z5}
	\end{centering}
\end{figure}

\clearpage
\begin{figure}[tb]
	\begin{centering}
		\includegraphics[width=0.7\textwidth]{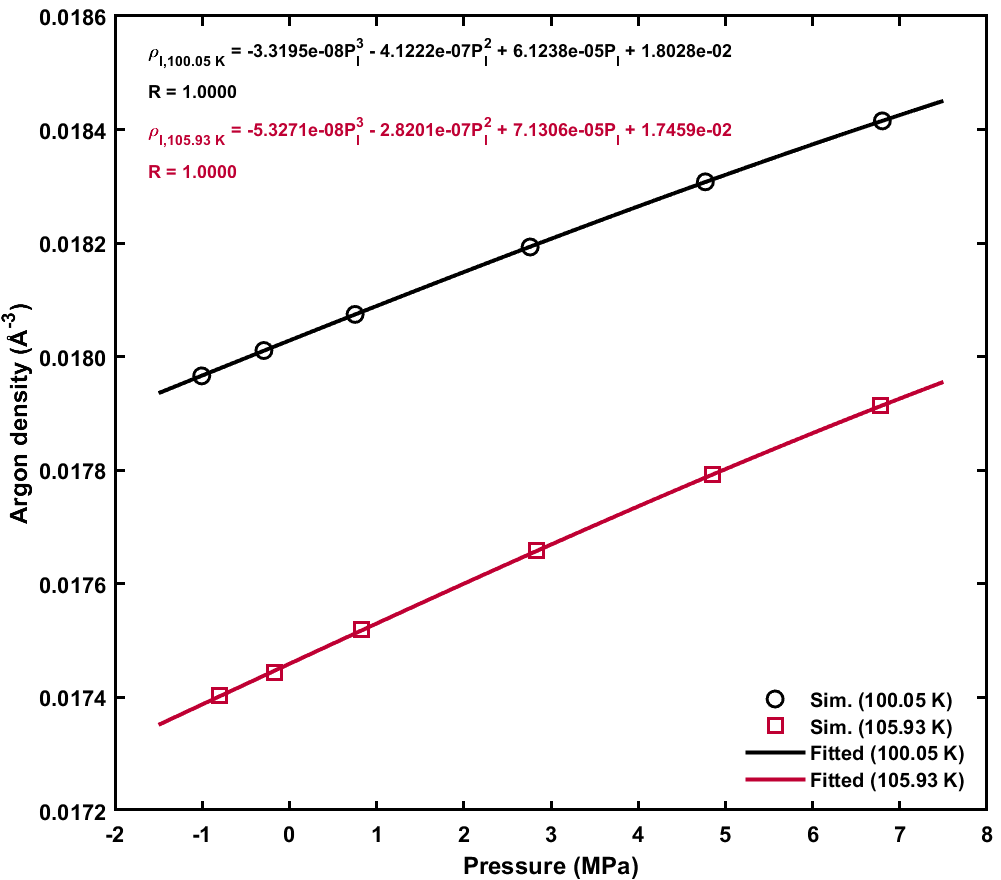}
		\caption{Liquid argon density as a function of pressure at 100.05 K and 105.93 K. The solid lines represent the fitted empirical EOSs. (This figure is intended for color reproduction on the Web and in print.)
		}
		\label{fig:z6}
	\end{centering}
\end{figure}

\clearpage
\begin{figure}[tb]
	\begin{centering}
		\includegraphics[width=0.9\textwidth]{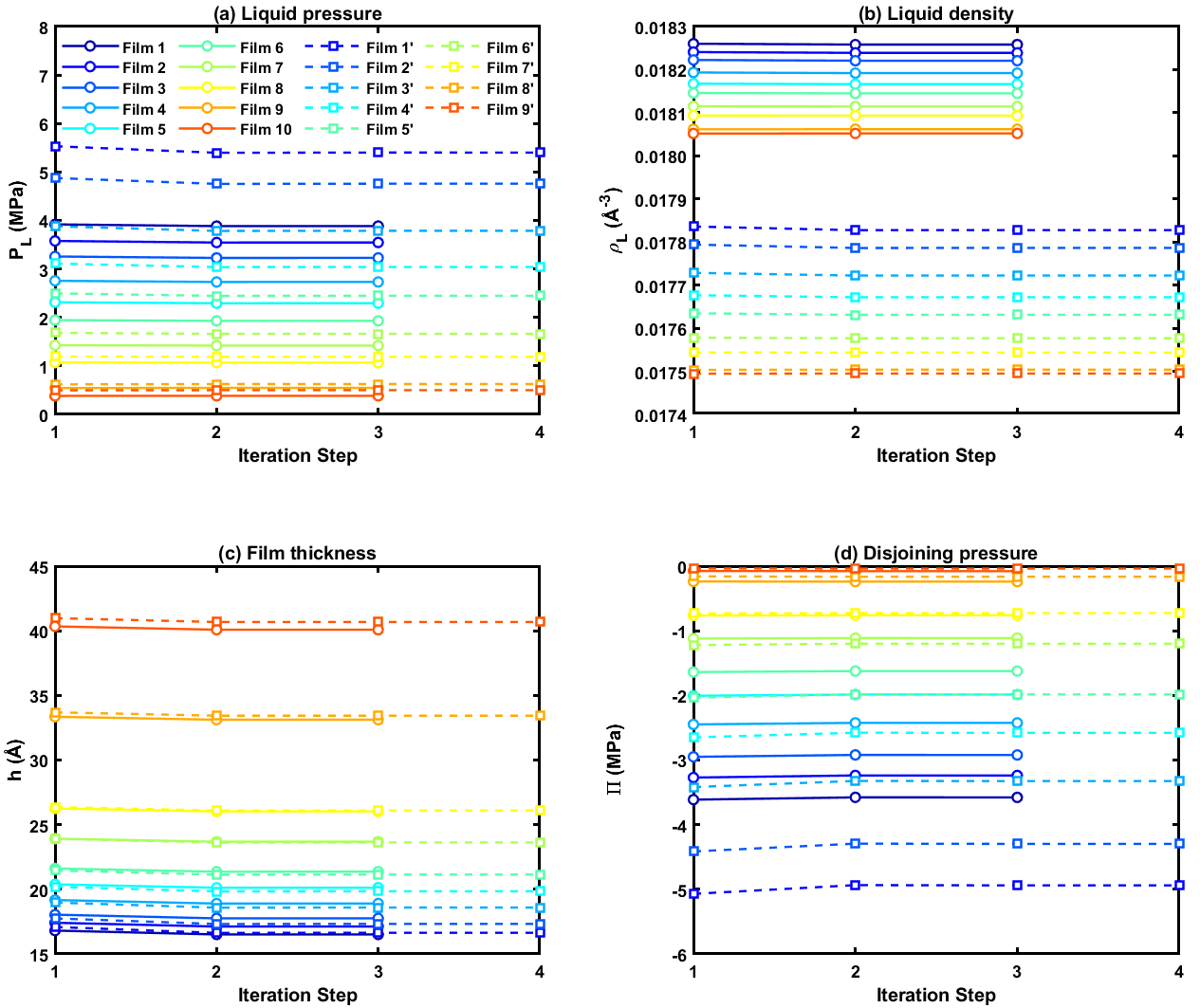}
		\caption{Convergence performance of the iteration algorithm for various properties of the argon nanofilm system: (a) liquid pressure, (b) liquid density, (c) film thickness, and (d) disjoining pressure as a function of iteration steps. The labels Films 1-10 and 1'-9' correspond to simulated film systems shown in Table 1 at 100.05 K and 105.93 K, respectively. (This figure is intended for color reproduction on the Web and in print.)
		}
		\label{fig:z7}
	\end{centering}
\end{figure}

\clearpage
\begin{figure}[tb]
	\begin{centering}
		\includegraphics[width=1.0\textwidth]{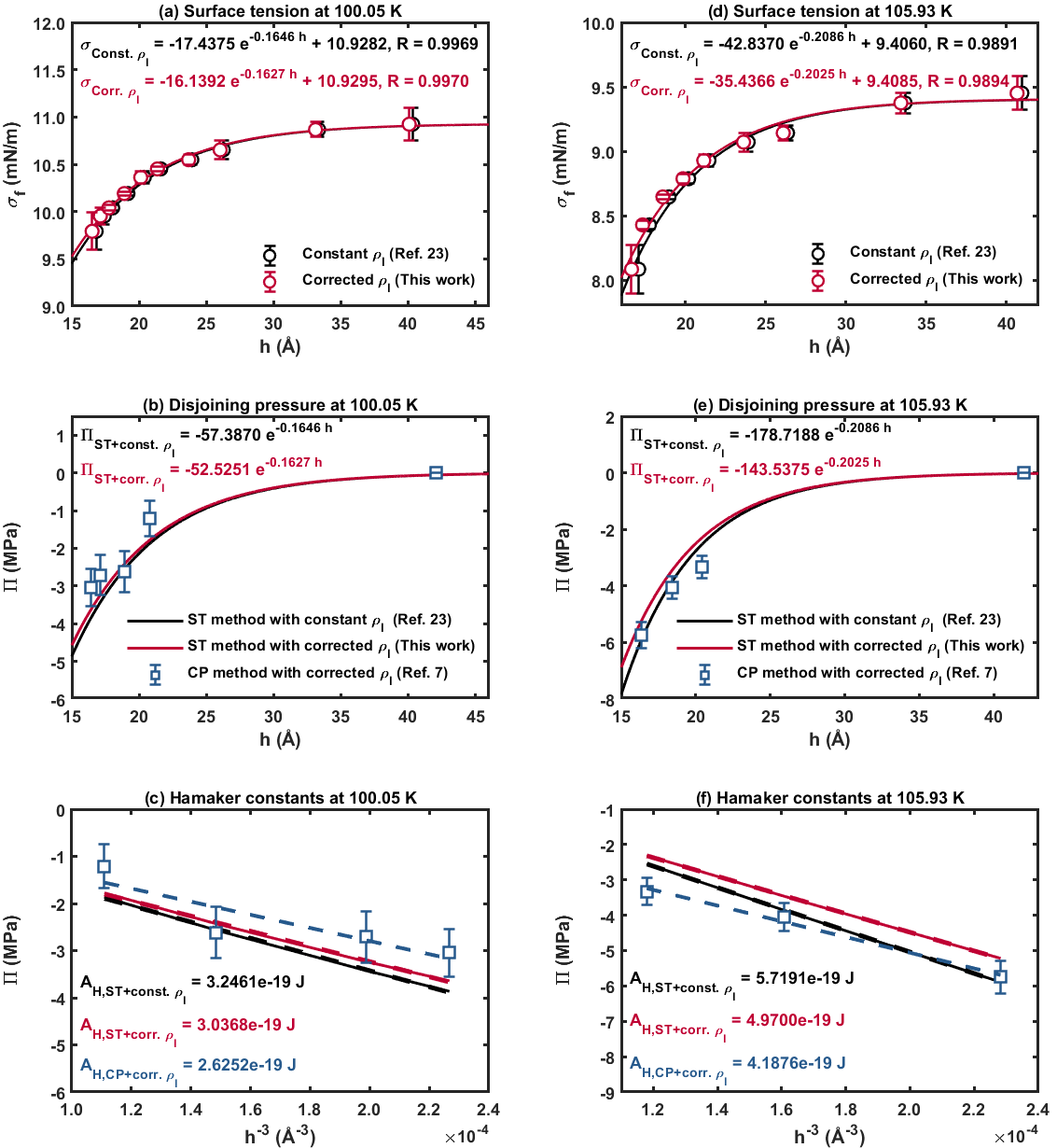}
		\caption{(a,d) Top panels: Surface tension of argon nanofilms as a function of film thickness; solid lines denote fitted curves.
			(b,e) Middle panels: Corresponding disjoining pressure as a function of film thickness.
			(c,f) Bottom panels: Corresponding disjoining pressure as a function of the inverse cubic film thickness; dashed lines indicate linear fits.
			The left and right columns correspond to temperatures of 100.05 K and 105.93 K, respectively.
			Our results are compared with those reported by Bhatt \textit{et al.} \cite{bhatt2002molecular} and Yang \textit{et al.} \cite{yang2026resolving}. (This figure is intended for color reproduction on the Web and in print.)
		}
		\label{fig:z8}
	\end{centering}
\end{figure}

\clearpage
\begin{figure}[tb]
	\begin{centering}
		\includegraphics[width=1.0\textwidth]{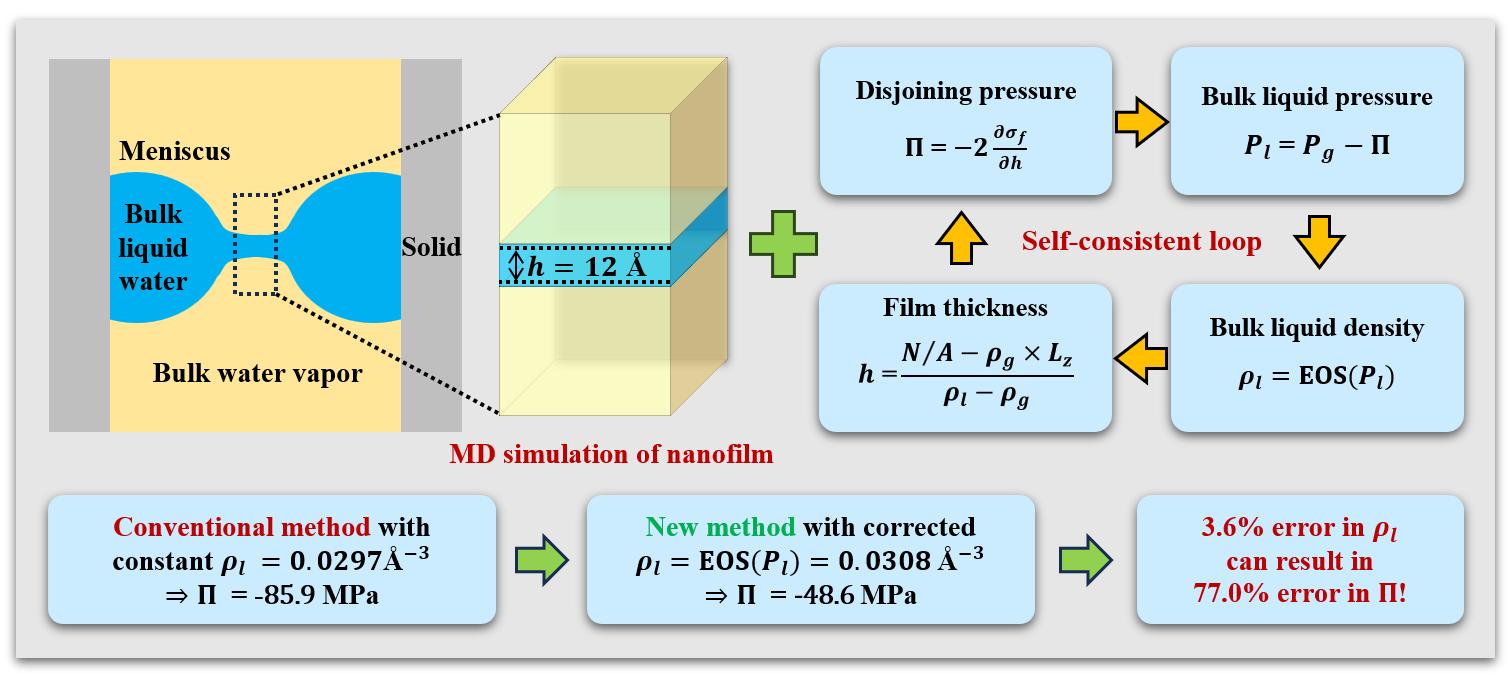}
		\caption{Schematic illustration of the self-consistent framework for determining the $\Pi$ of fluid nanofilms. Comparison with the conventional approach, which assumes a constant bulk liquid density of water, demonstrates that a modest 3.6\% error in $\rho_l$ can lead to a substantial 77.0\% error in the predicted $\Pi$ at $h$=12 \AA. (This figure is intended for color reproduction on the Web and in black-and-white in print.)
		}
		\label{fig:z9}
	\end{centering}
\end{figure}

\newpage
\begin{table}[!ht]
	\caption{Bulk liquid and interfacial properties at the initial ($k=0$) and final ($k=f$) iterations. The surface tension data are taken from Ref. \cite{yang2026resolving}.}
	\raggedright
	\begin{tabular}{c|c|c|c|c|c|c|c|c|c}
		\hline
		Sys. & ID & T \tiny K & $\sigma_f$ \tiny mN/m & $P_l^0$ \tiny MPa  & $P_l^f$ \tiny MPa & $\rho_l^0$ \tiny $\textup{\AA}^{-3}$ & $\rho_l^f$ \tiny $\textup{\AA}^{-3}$  & $h^0$ \tiny $\textup{\AA}$ & $h^f$ \tiny $\textup{\AA}$\\ \hline
		H$_2$O & 1 & 479.00  & 25.92  & 126.65  & 107.09  & 0.032126  & 0.031795  & 11.34  & 10.37  \\ 
		H$_2$O & 2 & 479.00  & 30.31  & 76.54  & 63.99  & 0.031274  & 0.031049  & 12.23  & 11.45  \\ 
		H$_2$O & 3 & 479.00  & 33.20  & 48.67  & 41.28  & 0.030759  & 0.030612  & 13.02  & 12.37  \\ 
		H$_2$O & 4 & 479.00  & 34.57  & 26.94  & 23.86  & 0.030308  & 0.030240  & 14.08  & 13.53  \\ 
		H$_2$O & 5 & 479.00  & 35.85  & 6.46  & 6.83  & 0.029827  & 0.029836  & 16.71  & 16.28  \\ 
		H$_2$O & 6 & 479.00  & 36.81  & 1.88  & 2.32  & 0.029710  & 0.029721  & 19.41  & 18.98  \\ 
		H$_2$O & 7 & 479.00  & 36.95  & 0.76  & 0.89  & 0.029681  & 0.029684  & 23.35  & 22.87  \\ 
		H$_2$O & 8 & 479.00  & 36.96  & 0.64  & 0.65  & 0.029678  & 0.029678  & 28.70  & 28.11  \\ 
		H$_2$O & 9 & 479.00  & 37.14  & 0.63  & 0.63  & 0.029677  & 0.029677  & 34.00  & 33.31  \\ 
		H$_2$O & 10 & 479.00  & 37.42  & 0.63  & 0.63  & 0.029677  & 0.029677  & 42.04  & 41.19  \\ \hline
		Ar & 1 & 100.05  & 9.79  & 3.92  & 3.88  & 0.018260  & 0.018258  & 16.81  & 16.51  \\ 
		Ar & 2 & 100.05  & 9.95  & 3.58  & 3.55  & 0.018241  & 0.018239  & 17.41  & 17.12  \\ 
		Ar & 3 & 100.05  & 10.04  & 3.26  & 3.23  & 0.018222  & 0.018221  & 18.04  & 17.75  \\ 
		Ar & 4 & 100.05  & 10.19  & 2.76  & 2.73  & 0.018193  & 0.018192  & 19.17  & 18.90  \\ 
		Ar & 5 & 100.05  & 10.36  & 2.31  & 2.29  & 0.018167  & 0.018166  & 20.38  & 20.12  \\ 
		Ar & 6 & 100.05  & 10.45  & 1.94  & 1.93  & 0.018146  & 0.018145  & 21.61  & 21.36  \\ 
		Ar & 7 & 100.05  & 10.55  & 1.43  & 1.42  & 0.018115  & 0.018115  & 23.92  & 23.68  \\ 
		Ar & 8 & 100.05  & 10.65  & 1.07  & 1.07  & 0.018093  & 0.018093  & 26.25  & 26.03  \\ 
		Ar & 9 & 100.05  & 10.87  & 0.54  & 0.55  & 0.018062  & 0.018062  & 33.35  & 33.12  \\ 
		Ar & 10 & 100.05  & 10.92  & 0.38  & 0.38  & 0.018052  & 0.018052  & 40.33  & 40.08  \\ 
		Ar & 1$'$ & 105.93  & 8.09  & 5.53  & 5.40  & 0.017836  & 0.017828  & 17.08  & 16.64  \\ 
		Ar & 2$'$ & 105.93  & 8.43  & 4.88  & 4.76  & 0.017794  & 0.017786  & 17.74  & 17.32  \\ 
		Ar & 3$'$ & 105.93  & 8.65  & 3.88  & 3.79  & 0.017728  & 0.017722  & 18.97  & 18.59  \\ 
		Ar & 4$'$ & 105.93  & 8.79  & 3.11  & 3.04  & 0.017677  & 0.017672  & 20.19  & 19.84  \\ 
		Ar & 5$'$ & 105.93  & 8.93  & 2.50  & 2.45  & 0.017635  & 0.017631  & 21.45  & 21.14  \\ 
		Ar & 6$'$ & 105.93  & 9.07  & 1.68  & 1.66  & 0.017578  & 0.017577  & 23.90  & 23.62  \\ 
		Ar & 7$'$ & 105.93  & 9.14  & 1.19  & 1.19  & 0.017544  & 0.017543  & 26.35  & 26.10  \\ 
		Ar & 8$'$ & 105.93  & 9.38  & 0.62  & 0.63  & 0.017503  & 0.017504  & 33.68  & 33.43  \\ 
		Ar & 9$'$ & 105.93  & 9.45  & 0.50  & 0.50  & 0.017494  & 0.017495  & 40.96  & 40.68 \\ \hline
	\end{tabular}
\end{table}

\end{document}